# Gas pressure manipulation of exciton states in monolayer WS$_2$

Shuangping Han[1], Pengyu Zan[1], Yu Yan[1], Yaoxing Bian[1], Chengbing Qin[1,2,*], Liantuan Xiao[1,2,*]

[1] College of Physics, Taiyuan University of Technology, Taiyuan, Shanxi 030006, China

[2] State Key Laboratory of Quantum Optics and Quantum Optics Devices, Institute of Laser Spectroscopy, Shanxi University, Taiyuan, Shanxi 030006, China

E-mail: chbqin@sxu.edu.cn, xlt@sxu.edu.cn

**Abstract**

Over the past few decades, thin film optoelectronic devices based on transition metal dichalcogenides (TMDs) have made significant progress. However, the sensitivity of the exciton states to environmental change presents challenges for device applications. This work reports on the evolution of photo-induced exciton states in monolayer WS$_2$ in a chamber with low gas pressure. It elucidates the physical mechanism of the transition between neutral and charged excitons. At 222 mTorr, the transition rate between excitons includes two components, 0.09 s$^{-1}$ and 1.68 s$^{-1}$, respectively. Based on this phenomenon, we have developed a pressure-tuning method for exciton manipulation, allowing a tuning range of approximately 40% in exciton weight. We also demonstrate that the intensity of neutral exciton emission from monolayer WS$_2$ follows a power-law distribution concerning gas pressure, indicating a highly sensitive pressure dependence. This work presents a non-destructive and highly sensitive method for exciton conversion through in-situ manipulation. It highlights the potential development of monolayer WS$_2$ in pressure sensing and explains the impact of environmental factors on product quality in photovoltaic devices.

## 1 Introduction

Two-dimensional (2D) materials, led by graphene, have undergone explosive development in recent years due to their unique electrical and optical properties and have attracted the interest of a large number of scientific researchers [1–6]. Among them, the development prospect of transition metal dichalcogenides (TMDs) is particularly concerned [7]. Monolayer TMDs are direct band gap semiconductors with strong Coulomb interaction, high photoluminescence (PL) intensity and broad PL spectra [1–3,7,8]. TMDs provide a new platform for studying complex many-body interactions, such as exciton-phonon transition, exciton-exciton annihilation, and exciton conversion, due to the exposure of 2D electrons to the external environment [7,9–13]. However, this feature is a double-edged sword, which brings excellent research values, but also faces some challenges [14,15]. Exciton lifetime has been increasingly taken into account with the use of ultrafast detector devices. The higher the content of short-lived neutral excitons, the faster the detection rate of TMDs-based photodetectors and the shorter the response time [16]. The emission of charged excitons indicates the charge doping of the material. The higher the proportion of negatively charged excitons, the more pronounced the n-doping of the material. The amount of negative charge doping is directly related to the sensitivity of the detection response in the field of device testing and photocurrent-based device development [16]. Therefore, it is necessary to realize the controllable exciton regulation of TMDs and analyze the variation of their optical properties within different environments [17–20].

The easy adhesion and large surface-to-volume ratio of monolayer TMDs make them extremely sensitive to the changes in their surrounding environments, such as gating [21], doping [22], defects [23], temperature [24], excitation power [25], and collection locations [26]. So far, abundant studies have been performed in these areas. The luminescence modulation of TMDs has been observed by laser irradiation, pressure regulation, charge doping, and other techniques. The intrinsic physical phenomena, such as exciton conversion, have been explained [22–24]. Liang et al. reversibly engineered the spin-orbit coupling of monolayer $MoS_2$ by laser irradiation under controlled gas environments, and the spin-orbit splitting has been effectively modified from 140 to 200 meV [27]. Furthermore, the PL intensity of the B exciton can be reversibly manipulated over 2 orders of magnitude. Yang et al. achieved the reversible engineering of neutral and charged excitons by switching the irradiation power, and attributed the results to laser-assisted adsorption and desorption of gas molecules [14]. Paradisanos et al. reported the temperature-dependent and power-dependent PL properties of $WS_2$. They proved that the neutral exciton and the charged exciton show a linear to sub-linear dependence over the entire temperature range, and determined that the binding energy of the biexciton is 65~70 meV [24]. Ma et al. realized the n-type and p-type doping of monolayer $WS_2$ by dropping the electron donor and acceptor solutions of triphenylphosphine and gold chloride on the surface of $WS_2$. The electrical properties of the prepared field effect transistor were studied, showing great potential for light detection [22]. Nayak's group investigated the structural, electronic, electrical, and vibrational properties of multilayer $WS_2$ at hydrostatic pressures up to 35 GPa. The results show that $WS_2$ undergoes an isostructural semiconductor-to-metallic transition at approximately 22 GPa at 280 K [15,28]. However, these exciton regulation methods have problems such as complex processes, high uncertainty, and poor stability. Therefore, in-situ non-destructive stable exciton engineering research is helpful in improving the response rate and detection sensitivity of thin-layer optoelectronic devices.

In this work, we propose an exciton states control method with high stability based on pressure-tuning. The PL evolution of monolayer $WS_2$ under low pressure proves that the exciton regulation by pressure is more adjustable and stable than in ambient conditions. We show that the environmental evolution of the different excitons is due to gas desorption caused by pressure changes, and we obtain the underlying physical mechanisms and the exciton-exciton transition rates. This method can provide ~40% adjustable range of exciton-exciton conversion efficiency. The pressure as a function of the exciton intensity obeys the power law distribution with the power exponent ~2.38. Moreover, we verify that the change of the dissociation energy of trion with gas pressure originates from the variation of binding mode as the concentration of two-dimensional electron gas (2DEG) evolved. This work realizes the stable control technology of monolayer $WS_2$ exciton, which can promote the development and application of thin-layer optoelectronic devices in extreme environments, and contribute to the promotion of intelligent manufacturing industry and biosensors.

## 2 Experimental

The PL emission spectra of monolayer $WS_2$ are performed on a homemade confocal fluorescence imaging system, and the experimental setup is illustrated in Fig. 1a. A pulsed laser with a wavelength of 532 nm (Toptica) is employed to excite the sample and conduct PL imaging. The laser goes through a beam expander, filter, X-Y scanner, and 4f system before focusing on the sample by a dry objective lens (100×, NA=0.8), ensuring precise optical manipulation. $WS_2$ is positioned within a vacuum chamber to control the pressure of the surrounding environment. A beam splitter (BS) is used to separate the PL signal for subsequent spectral and intensity detection. A monochromator (Andor Shamrock SR-303i) and a CCD camera (Andor DR-316B-LDC-DD) are jointly utilized to generate the spectra. Additionally, the time-dependent emission intensity is recorded using the single photon detection module (SPCM). Thus, the tunability of $WS_2$ PL spectra enabled by gas pressure can be recorded and analyzed. In the subsequent experimental study, all tests were completed at room temperature.

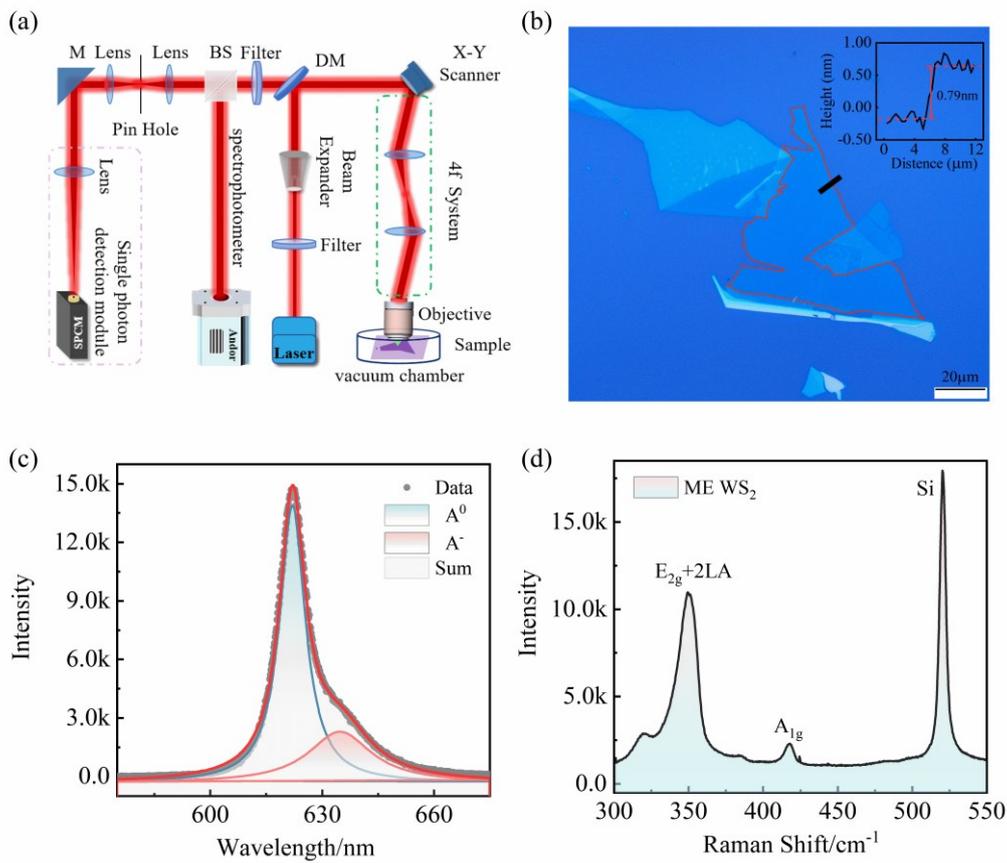

**Fig. 1** Experimental setup and sample characterization. (a) Schematic diagram of the confocal microscopy imaging system. The 4f system ensures the light focuses on the sample within the scanning range. M: mirror; DM: dichroic mirror; SPCM: Single photon detection module; (b) Optical imaging of the mechanically exfoliated (ME) $WS_2$. The contour surrounded by the red dotted line is the monolayer $WS_2$ region. The inset is the height profile of the marked line in the optical imaging through atomic force microscopy (AFM). Scale bar: 20 μm; (c) PL spectra of ME 1L-$WS_2$. The gray dots represent the experimental data. The solid red line is the sum curve obtained through the 2-peak Lorentz fitting. The green and magenta represent the emission of $A^0$ and $A^-$ excitons; (d) Raman spectra of ME 1L-$WS_2$. The Raman signal of silicon substrate is located at 521 cm$^{-1}$. The other two peaks are E$_{2g}$ and A$_{1g}$ Raman modes of the WS$_2$, with 2LA emission also included in the left peak. Test temperature: room temperature.

# 3 Results and discussion

## 3.1 Sample characterization

The layered WS$_2$ used in the experiment is prepared by mechanical exfoliation (ME) from bulk materials and deposited on SiO$_2$/Si substrates. Typical sample characterizations are shown in Fig. 1. The optical imaging in Fig. 1b shows good sample homogeneity, with an area larger than 150 μm$^2$. The inset shows the morphology of WS$_2$ measured by atomic force microscopy (AFM), with a sample thickness of approximately 0.8 nm, confirming that the thin film is a monolayer. Figs. 1c and 1d show the PL and Raman spectra performed under atmospheric pressure. The asymmetric PL spectrum is actually caused by the joint emission of A$^0$ and A$^-$. Through the Lorentz bimodal fitting, we can demonstrate the features of these two peaks well. The high-energy peak (615 nm) is generated by the neutral exciton emission, while the low-energy peak is attributed to the charged exciton emission caused by the additional charge present in WS$_2$, indicating a high density of extra electrons in this layer. This phenomenon generally comes from two aspects. The first one is the impurities introduced during crystal growth, which are more commonly found on samples prepared by chemical vapor deposition. The other one is the charge adsorption caused by long-term exposure of crystals to the external environment. It is mainly due to the increase of surface charge induced by the interaction between the material surface and the electrons/ions in the surrounding atmosphere. The research object of this study is ME monolayer-WS$_2$ (1L-WS$_2$), which suggests a strong possibility of originating from the second point. For the convenience of subsequent description and understanding, we have defined spectral weights as the proportion of each exciton emission intensity to the total emission spectral intensity, that is, $W(A^0,t) = I_{A^0} / (I_{A^0} + I_{A^-})$, $W(A^-,t) = I_{A^-} / (I_{A^0} + I_{A^-})$. Fig. 1d shows the Raman signal of the sample. The Raman signal at 521 cm$^{-1}$ is a typical signal from the silicon substrate. Interlayer vibration (A$_{1g}$) is located at 418 cm$^{-1}$. The signal at 350 cm$^{-1}$ comes from the combined effect of in-plane vibration (E$_{2g}$) and second-order Raman resonance of longitudinal acoustic phonons (2LA), which is also why it appears to be composed of two peaks. This phenomenon is particularly evident when excited by a 532 nm laser [9]. The wavenumber difference between the two peaks of ~68 cm$^{-1}$ further confirms that the sample is a monolayer.

## 3.2 WS$_2$ PL spectra

In previous reports, the methods to realize the regulation of exciton emission in monolayer 2D materials include doping, laser irradiation, etc., which have the shortcomings of complicated operation and high uncertainty. To solve these problems and realize the stable engineering of exciton emission, this work proposes to achieve this goal by manipulating the ambient pressure. We first measured PL spectra at atmospheric pressure and high vacuum (222 mTorr). Figs. 2a and 2c present the 2D intensity maps, respectively. The abscissa represents the wavelength, and the ordinate is the time of laser irradiation (the spectral acquisition integration time is 0.1 s/frame). The value of each point represents the PL intensity of the sample at that wavelength. Correspondingly, the trajectories in Figs. 2b and 2d represent the time evolution of PL emission corresponding to the two vertical lines (λ=615 nm, 630 nm, corresponding to the neutral exciton and charged exciton emission of WS$_2$, respectively) in Figs. 2a and 2c. We can demonstrate very significant differences in intensity and spectral changes. The PL intensity shows a sharp decline in both environments, but the modulation under the atmospheric pressure is significantly faster, and the intensity basically reaches a stable stage within 2 s. The spectral weight of A⁻ is small, and the change is not obvious, which is not easy to observe. In addition, the emission characteristics of WS$_2$ at atmospheric pressure are closely related to the surrounding environment, so the stability is poor. The PL spectrum at low pressure shows obvious bimodal emission from the beginning of irradiation, and both the spectral weight and intensity are larger and more stable.

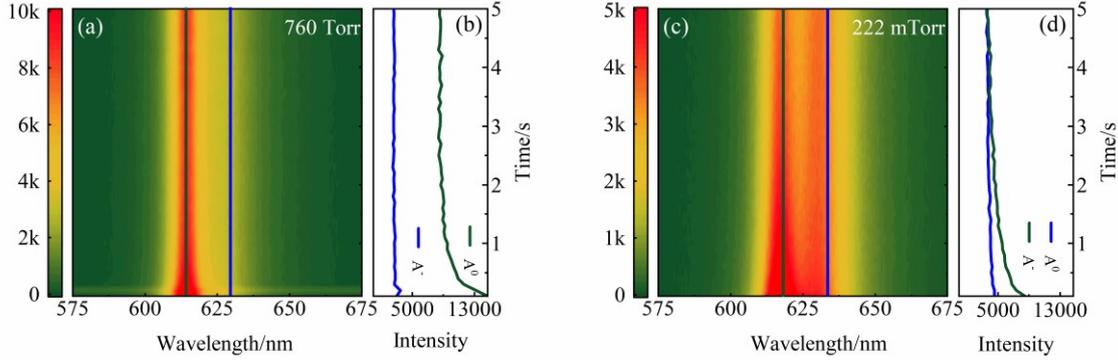

**Fig. 2** Characterization of PL spectra under 760 Torr and 222 mTorr. (a, b) 2D-intensity imaging of ME 1L-WS$_2$ under 760 Torr and 222 mTorr, respectively. The horizontal axis is the wavelength, while the vertical axis represents time (determined by the number of acquisition frames. Frame rate: 0.1 s/frame); (c, d) Time evolution of PL intensity at the peak position of A$^0$ and A$^-$ under 760 Torr and 222 mTorr.

The effect of pressure on the emission spectra of 2D materials was mainly analyzed from the evolution of band structures. However, the band gap variation only occurs when the external pressure changes significantly. Thus, the minor pressure change has not been paid attention to before. It is worth noting that the PL spectra of monolayer TMDs exhibit astonishing sensitivity to ambient gas atmospheres, which means that environmental variations (especially gas concentration) caused by pressure changes cannot be ignored, as seen from our experimental results. To further understand the causes, we will carefully discuss the evolution of PL behavior under low pressure, including the changes in intensity and spectral components, and clarify the internal physical mechanism through global fitting.

3.3 Exciton regulation mechanism at low pressure

To explain the above phenomenon, we carefully analyzed the situation at 222 mTorr. Firstly, the repeatability of the phenomenon can be seen in Fig. 3a, where different points exhibit high consistency after illumination, which proves the excellent homogeneity of the sample. Subsequently, we present the PL spectra of point 1 at several times in Fig. 3b. Except for common phenomena such as significant intensity reduction and two-peak emission, the decline of the two peaks is generally consistent with previous reports [29]. The upper panel of Fig. 3c is the function of neutral exciton-charged exciton conversion as a function of the observation time. The longitudinal coordinate is the spectral weight of A$^0$ and A$^-$ after the normalization of total spectral intensity. The two curves intersect at t = 0.57 s, followed by a reversal of the proportion of the two exciton emissions, which depends on factors such as irradiation power and the initial state of the sample.

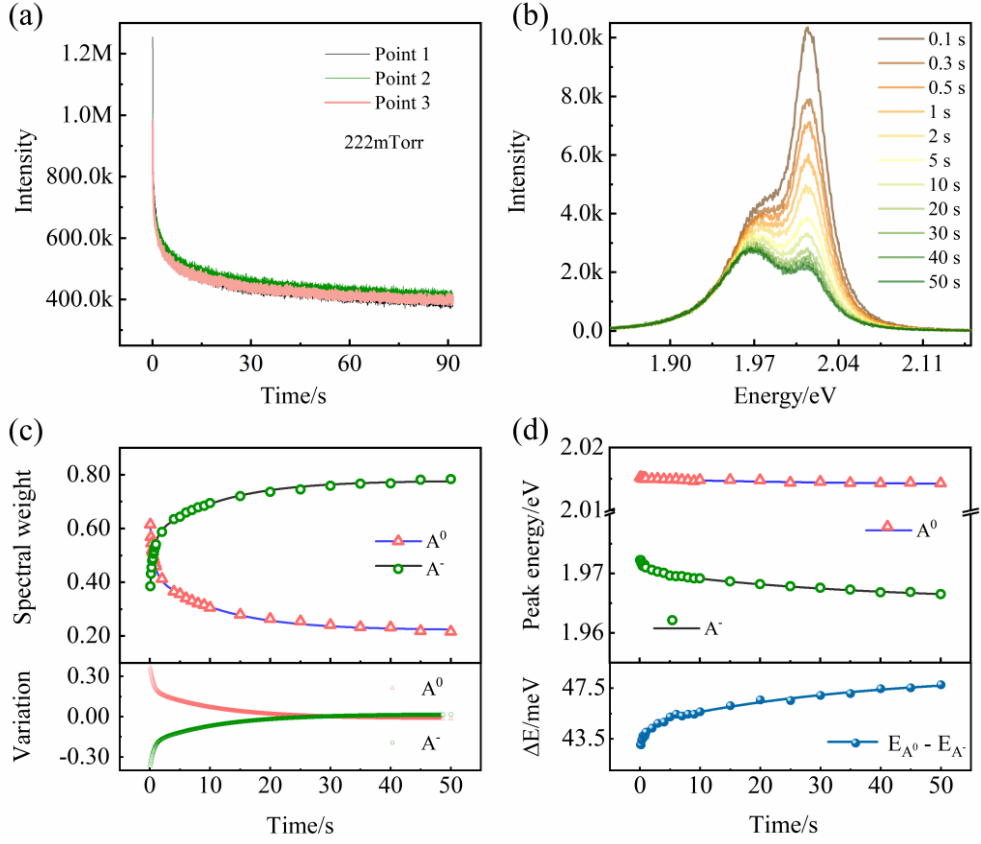

**Fig. 3** Emission spectra of ME 1L-WS$_2$ at 222 mTorr. (a) PL intensity as a function of irradiation time; (b) PL spectra of ME 1L-WS$_2$ at several times under 222 mTorr; (c) Upper panel: Evolution of neutral exciton-charged exciton conversion with time, where the magenta triangle and green circle are the spectral weights of A$^0$ and A$^-$, respectively. The blue and black solid lines are the fitting results, which follow a biexponential distribution, representing the roles of laser radiation and gas pressure in the conversion process, respectively. Lower panel: The variation of the spectral weight of each exciton relative to the stable-stage average value; (d) Upper panel: The time evolution of the peak energy of A$^0$ and A$^-$ with time by Lorentz bimodal fitting. Lower panel: The variation in peak energy difference between A$^0$ and A$^-$ over time.

Because of the limited dimension and large specific surface area, layered 2D materials readily absorb gas, and their emission spectra are easily modified by the surrounding environment (especially charged ions in the air) [26]. The essence of this process is the change of a 2D electron gas (2DEG) concentration induced by the gas desorption on the sample surface. The theoretical analysis of the related phenomena caused by laser irradiation has been reported in our previous work [14]. When the irradiation power reaches the activation energy of the desorption gas molecules, the adsorbed gas molecules will leave the surface of the monolayer WS$_2$, and the intensity decay of PL shows a monoexponential distribution. The decay curve here follows a bi-exponential distribution, indicating that gas pressure also plays a crucial role in addition to laser irradiation. Although weak adsorption may also occur under low pressure, we can still imagine that the number of desorbed molecules is much larger than that of adsorbed, so the adsorption effect can be ignored. The following formula can be used to simulate the exciton-exciton conversion of A$^0$ and A$^-$ with time:

$$W(t) = W(0) + a_1 \cdot \exp(k_1 \times t_1) + a_2 \cdot \exp(k_2 \times t_2) \tag{1}$$

Here, W(t) is the spectral weight of $A^0$ or $A^-$ at time t, and k is the conversation rate of $A^0$ ($A^-$) to $A^-$ ($A^0$). According to the fitting curve (as the solid line shown in Fig. 3c), $k_1$ and $k_2$ are determined to be 0.089 s$^{-1}$ and 1.675 s$^{-1}$, respectively. The former is basically consistent with the laser-induced exciton conversion rate reported in other work [14], while the latter is attributed to pressure induction, which occurs in a shorter time. In the lower panel of Fig. 3c, we show the variation with the spectral weight of the stable stage as the average (0.24 for $A^0$ and 0.76 for $A^-$, respectively). From 20 s, the variation is less than 10$^{-3}$ and can be maintained for a long time, which further proves the stability of exciton engineering at low pressure.

**Table 1** Bi-exponential fitting results of the curves in Figures 3c and 3d, in which the conversion rate of $A^-$ and $\Delta E$ shows a high consistency.

|  | W(t) | $A^0$ | $A^-$ | $\Delta E$ |
|---|---|---|---|---|
| $k_1$ (s$^{-1}$) | 0.089 | 0.027 | 0.029 | 0.029 |
| $k_2$ (s$^{-1}$) | 1.675 | 2.146 | 0.828 | 0.733 |

Furthermore, we noticed that during this process, the peak position of $A^0$ shifts from 2.015 eV to 2.014 eV, while $A^-$ shifts from 1.972 eV to 1.966 eV. The energy of the $A^0$ peak is slightly weakened, while the energy of the $A^-$ peak is reduced more, suggesting an efficient energy transfer from higher to lower energy, which is also the source of PL quenching. Their energy difference ($\Delta E = E_{A0} - E_{A-}$), that is, the dissociation energy (or binding energy) of the charged exciton, increased from 43.0 meV to 44.8 meV (Fig. 3d), due to the change in the concentration of 2DEG. The decrease of the adsorbed molecule concentration during the quenching process increases the free electrons, and the neutral excitons are more likely to combine with the high-energy spin states to form charged excitons at high 2DEG concentrations [30]. Although their variations are relatively weak, the global fitting shows a bi-exponential behavior similar to the trend of spectral weight over time, and the fitting results are shown in Table 1. Therefore, we can conclude that the PL of 1L-WS$_2$ is modified by both high vacuum (low pressure) and laser irradiation. The rate obtained by curve fitting of $A^-$ and $\Delta E$ is basically the same, which indicates the correctness of the above theoretical analysis.

### 3.4 Exciton engineering based on pressure regulation

After an in-depth understanding of the underlying mechanism at low pressure, we found that compared with the case of pure laser irradiation, the emission of two excitons in the PL spectra of WS$_2$ at low pressure is more pronounced, and the different positions show a high degree of consistency. This result indicates that low pressure can effectively regulate exciton emission. In the following work, we realized the steady-state exciton engineering by adjusting the pressure. The emission intensity of monolayer WS$_2$ is rapidly quenched with laser irradiation and reaches a stable state within less than 5 seconds, consistent with the phenomenon at 222 mTorr. To clarify the effect of pressure on the conversion between different excitons, we define the increase of neutral excitons ($\eta_{A0}$) or the decrease of charged excitons ($\eta_{A-}$) as the exciton conversion efficiency, that is:

$$\eta_{A0} = \left(\frac{I_{A0}}{I}\right)_{P_2} - \left(\frac{I_{A0}}{I}\right)_{P_1} = W(P_{2,A0}) - W(P_{1,A0}) \qquad (2)$$

$$\eta_{A-} = \left(\frac{I_{A-}}{I}\right)_{P_2} - \left(\frac{I_{A-}}{I}\right)_{P_1} = W(P_{2,A-}) - W(P_{1,A-}) \qquad (3)$$

Where $I$ is the total emission intensity of WS$_2$, $I_{A^0}$, $I_{A^-}$ is the emission intensity of neutral excitons and charged excitons obtained by Lorentz bimodal fitting; $P_1$ and $P_2$ represent different pressures. $W(P_{2,A0})$ represents the spectral weight of A$^0$ at pressure $P_2$. Similarly, the physical meanings of $W(P_{1,A0})$, $W(P_{2,A-})$ and $W(P_{1,A-})$ can also be clarified. $\eta_{A0}$ and $\eta_{A-}$ represent the efficiency of the conversion from A$^0$(A$^-$) to A$^-$(A$^0$). As the neutral exciton increases, the charged exciton decreases proportionally, i.e., $|\eta_{A0}| = |\eta_{A-}|$. The conversion efficiency between A$^0$ and A$^-$ exciton is as high as 40% at atmospheric pressure (as shown in Fig. 4a), which provides a wide range of adjustment redundancy.

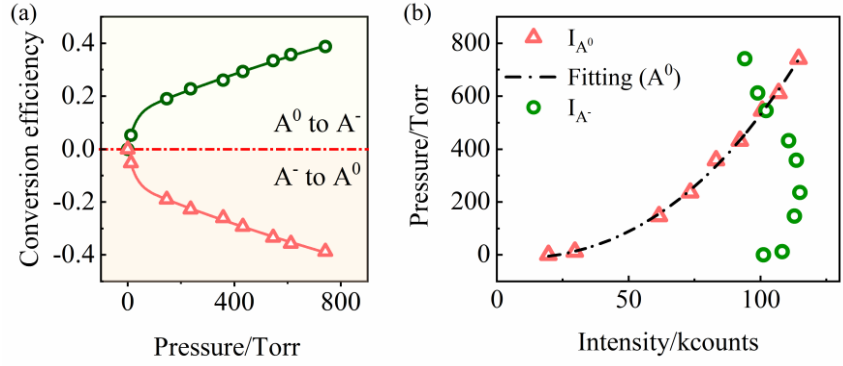

**Fig. 4** Exciton engineering based on pressure. (a) The evolution of conversion efficiency between A$^0$ and A$^-$ with pressure. The green circle and red triangle are the fitting results of $\eta_{A0}$ and $\eta_{A-}$, respectively. (b) The evolution of the pressure as a function of the emission intensity. The red triangles correspond to A$^0$; the dotted line is the power-law fitting trajectory. The green circle represents the charged exciton A$^-$.

Meanwhile, the peak center of A$^0$ is shifted from 2.016 eV to 2.015 eV, and the position of A$^-$ is increased from 1.968 eV to 1.975 eV, which proves that the quenching of PL comes from the conversion between A$^0$ and A$^-$. The dissociation energy of the charged exciton decreases from 48.6 meV to 40.1 meV, resulting from the decrease of 2DEG concentration with pressure. In this case, neutral excitons are more likely to combine with low-energy spin states to form charged excitons. In addition, the relationship between the pressure and the stable-stage PL intensity is shown in Fig. 4b. The red triangle represents the strength of A$^0$ measured experimentally under several different pressures, and the green circle represents A$^-$. The pressure as the function of the intensity for the A$^0$ peak obeys the power-law distribution:

$$P = c + A \times I^{\alpha} \tag{4}$$

Where $P$ and $I$ represent the pressure and emission intensity, respectively. $c$ is a constant term. $A$ is the transformation coefficient between pressure and the intensity of A$^0$. The power exponent α=2.38, indicating that the neutral exciton state is sensitive to pressure changes. There is a relationship $\log P \propto \alpha \times I$ due to the gradual increase in the number of gas molecules adsorbed on the sample's surface as the pressure increases. From this formula, it can be inferred that $\log P$ has a linear relationship with the emission intensity of neutral excitons, and the slope represents α. The larger the value of α, the more sensitive the response of WS$_2$ to pressure changes. The results show that the neutral exciton of monolayer WS$_2$ has good pressure sensitivity, while A$^-$ exhibits poor sensitivity. This characteristic of A$^0$ is expected to be used to develop ultra-thin wearable pressure sensors.

## 4 Conclusions

In view of the critical obstacles faced by the current photoelectric detection industry in terms of detection sensitivity and response rate, the research content of this work focuses on precise exciton regulation. Compared with the case of laser-only, the exciton adjustment at low pressure is more stable and easier to control. The modulation of PL intensity is attributed to the energy transfer between $A^0$ and $A^-$, due to the influence of gas pressure. The evolution of the spectral weights of $A^0$ and $A^-$ emission over time satisfies the two-exponential distribution, indicating that the exciton conversion rates caused by light irradiation and gas pressure are 0.09 $s^{-1}$ and 1.68 $s^{-1}$, respectively. The energy difference between the two excitons, *i.e.,* the dissociation energy of the charged exciton, gradually increases with the irradiation time. This is due to the decrease of the concentration of adsorbed molecules during the quenching process, and the increase of free electrons on the surface of the sample, leading to the rise of 2DEG concentration. At high 2DEG concentrations, neutral excitons are more likely to combine with high-energy spin states to form charged excitons, and the binding energy increases. In addition, by adjusting the surrounding pressure, the stable regulation of neutral excitons and charged excitons is realized, and the tunable range of conversion efficiency between $A^0$ and $A^-$ is ~40%. Finally, the analysis shows that the pressure and the emission intensity of $A^0$ state in monolayer $WS_2$ follow a power-law distribution, which indicates that $WS_2$ has good pressure sensitivity. Based on optical in-situ manipulation, our work provides a non-destructive and high-stability exciton regulation method. The results provide a theoretical basis for the manufacture and performance manipulation of thin-layer optoelectronic devices, and offer a new idea for the preparation and promotion of pressure-sensitive sensors and micro-biochemical sensors.

**Competing interests** The authors declare that they have no competing interests.

**Acknowledgments** The authors gratefully acknowledge support from the National Key Research and Development Program of China (Grant No. 2022YFA1404201), National Natural Science Foundation of China (Nos. U23A20380, U22A2091, 62222509, 62127817, and 6191101445), Shanxi Province Science and Technology Innovation Talent Team (No. 202204051001014), 111 projects (Grant No. D18001) and Shanxi Provincial Basic Research Program Project (202203021222107, 202203021222133).

## References

1. Chaves A, Azadani J G, Alsalman H, da Costa D R, Frisenda R, Chaves A J, Song S H, Kim Y D, He D, Zhou J, et al. Bandgap engineering of two-dimensional semiconductor materials. npj 2D Materials and Applications, 2020, 4(1): 1-21.
2. Cui C, Xue F, Hu W J, Li L J. Two-dimensional materials with piezoelectric and ferroelectric functionalities. npj 2D Materials and Applications, 2018, 2(1): 1-14.
3. Peng Z, Chen X, Fan Y, Srolovitz D J, Lei D. Strain engineering of 2D semiconductors and graphene: from strain fields to band-structure tuning and photonic applications. Light: Science & Applications, 2020, 9(1): 190.
4. Naik M H, Regan E C, Zhang Z, Chan Y H, Li Z, Wang D, Yoon Y, Ong C S, Zhao W, Zhao S, et al. Intralayer charge-transfer moiré excitons in van der Waals superlattices. Nature, 2022, 609(7925): 52-57.
5. Su R, Kuiri M, Watanabe K, Taniguchi T, Folk J. Superconductivity in twisted double bilayer graphene stabilized by $WSe_2$. Nature Materials, 2023, 22(11): 1332-1337.
6. Wang Y, Seki T, Yu X, Yu C C, Chiang K Y, Domke K F, Hunger J, Chen Y, Nagata Y, Bonn M. Chemistry governs water organization at a graphene electrode. Nature, 2023, 615(7950): E1-E2.
7. Schneider C, Glazov M M, Korn T, Höfling S, Urbaszek B. Two-dimensional semiconductors in the regime of strong light-matter coupling. Nature Communications, 2018, 9(1): 2695.

8. Qin C, Gao Y, Qiao Z, Xiao L, Jia S. Atomic-Layered MoS$_2$ as a Tunable Optical Platform. Advanced Optical Materials, 2016, 4(10): 1429-1456.

9. Han S, Boguschewski C, Gao Y, Xiao L, Zhu J, Loosdrecht P H M van. Incoherent phonon population and exciton-exciton annihilation dynamics in monolayer WS$_2$ revealed by time-resolved Resonance Raman scattering. Optics Express, 2019, 27(21): 29949-29961.

10. Chernikov A, van der Zande A M, Hill H M, Rigosi A F, Velauthapillai A, Hone J, Heinz T F. Electrical Tuning of Exciton Binding Energies in Monolayer WS$_2$. Physical Review Letters, 2015, 115(12): 126802.

11. Guo L, Chen C A, Zhang Z, M. Monahan D, Lee Y H, R. Fleming G. Lineshape characterization of excitons in monolayer WS$_2$ by two-dimensional electronic spectroscopy. Nanoscale Advances, 2020, 2(6): 2333-2338.

12. Gelly R J, Renaud D, Liao X, Pingault B, Bogdanovic S, Scuri G, Watanabe K, Taniguchi T, Urbaszek B, Park H, et al. Probing dark exciton navigation through a local strain landscape in a WSe$_2$ monolayer. Nature Communications, 2022, 13(1): 232.

13. Han S, Liang X, Qin C, Gao Y, Song Y, Wang S, Su X, Zhang G, Chen R, Hu J, et al. Criteria for Assessing the Interlayer Coupling of van der Waals Heterostructures Using Ultrafast Pump–Probe Photoluminescence Spectroscopy. ACS Nano, 2021, 15(8): 12966-12974.

14. Yang C, Gao Y, Qin C, Liang X, Han S, Zhang G, Chen R, Hu J, Xiao L, Jia S. All-Optical Reversible Manipulation of Exciton and Trion Emissions in Monolayer WS$_2$. Nanomaterials, 2020, 10(1): 23.

15. Nayak A P, Yuan Z, Cao B, Liu J, Wu J, Moran S T, Li T, Akinwande D, Jin C, Lin J F. Pressure-Modulated Conductivity, Carrier Density, and Mobility of Multilayered Tungsten Disulfide. ACS Nano, 2015, 9(9): 9117-9123.

16. Sharma A, Zhu Y, Halbich R, Sun X, Zhang L, Wang B, Lu Y. Engineering the Dynamics and Transport of Excitons, Trions, and Biexcitons in Monolayer WS$_2$. ACS Applied Materials & Interfaces, 2022, 14(36): 41165-41177.

17. Conti S, Pimpolari L, Calabrese G, Worsley R, Majee S, Polyushkin D K, Paur M, Pace S, Keum D H, Fabbri F, et al. Low-voltage 2D materials-based printed field-effect transistors for integrated digital and analog electronics on paper. Nature Communications, 2020, 11(1): 3566.

18. Lien D H, Amani M, Desai S B, Ahn G H, Han K, He J H, Ager J W, Wu M C, Javey A. Large-area and bright pulsed electroluminescence in monolayer semiconductors. Nature Communications, 2018, 9(1): 1229.

19. Sajid M, Osman A, Siddiqui G U, Kim H B, Kim S W, Ko J B, Lim Y K, Choi K H. All-printed highly sensitive 2D MoS$_2$ based multi-reagent immunosensor for smartphone based point-of-care diagnosis. Scientific Reports, 2017, 7(1): 5802.

20. Yu Y, Fong P W K, Wang S, Surya C. Fabrication of WS$_2$/GaN p-n Junction by Wafer-Scale WS$_2$ Thin Film Transfer. Scientific Reports, 2016, 6(1): 37833.

21. Tagarelli F, Lopriore E, Erkensten D, Perea-Causín R, Brem S, Hagel J, Sun Z, Pasquale G, Watanabe K, Taniguchi T, et al. Electrical control of hybrid exciton transport in a van der Waals heterostructure. Nature Photonics, 2023, 17(7): 615-621.

22. Ma X, Zhang R, An C, Wu S, Hu X, Liu J. Efficient doping modulation of monolayer WS$_2$ for optoelectronic applications. Chinese Physics B, 2019, 28(3): 037803.

23. Chow P K, Jacobs-Gedrim R B, Gao J, Lu T M, Yu B, Terrones H, Koratkar N. Defect-Induced Photoluminescence in Monolayer Semiconducting Transition Metal Dichalcogenides. ACS Nano, 2015, 9(2): 1520-1527.

24. Paradisanos I, Germanis S, Pelekanos N T, Fotakis C, Kymakis E, Kioseoglou G, Stratakis E. Room temperature observation of biexcitons in exfoliated WS$_2$ monolayers. Applied Physics Letters, 2017, 110(19), 193102.

25. Zhu B, Chen X, Cui X. Exciton Binding Energy of Monolayer WS$_2$. Scientific Reports, 2015, 5(1): 9218.

26. Kim M S, Yun S J, Lee Y, Seo C, Han G H, Kim K K, Lee Y H, Kim J. Biexciton Emission from Edges and Grain Boundaries of Triangular WS$_2$ Monolayers. ACS Nano, 2016, 10(2): 2399-2405.

27. Liang X, Qin C, Gao Y, Han S, Zhang G, Chen R, Hu J, Xiao L, Jia S. Reversible engineering of spin–orbit splitting in monolayer MoS$_2$ via laser irradiation under controlled gas atmospheres. Nanoscale, 2021, 13(19): 8966-8975.

28. Li L, Zeng Z Y, Liang T, Tang M, Cheng Y. Elastic Properties and Electronic Structure of $WS_2$ under Pressure from First-principles Calculations. Zeitschrift für Naturforschung A, 2017, 72(4): 295-301.
29. Xu S, Sun J, Weng L, Hua Y, Liu W, Neville A, Hu M, Gao X. In-situ friction and wear responses of $WS_2$ films to space environment: Vacuum and atomic oxygen. Applied Surface Science, 2018, 447: 368-373.
30. Jadczak J, Kutrowska-Girzycka J, Kapuściński P, Huang Y S, Wójs A, Bryja L. Probing of free and localized excitons and trions in atomically thin $WSe_2$, $WS_2$, $MoSe_2$ and $MoS_2$ in photoluminescence and reflectivity experiments. Nanotechnology, 2017, 28(39): 395702.